\documentclass[aip,jap,amsmath,amssymb,
reprint
]{revtex4-1}

\usepackage{graphicx}
\usepackage{hyperref}
\makeatletter
\def\boldit#1{{\reset@font\bfseries\itshape#1}}
\makeatother

\draft % marks overfull lines with a black rule on the right

\begin{document}

% Use the \preprint command to place your local institutional report number 
% on the title page in preprint mode.
% Multiple \preprint commands are allowed.
\preprint{Abstract: No. DX-09}

\title{\boldit{Ab initio} study of magnetic anisotropy in cobalt doped zinc oxide with electron-filling} %Title of paper

% repeat the \author .. \affiliation  etc. as needed
% \email, \thanks, \homepage, \altaffiliation all apply to the current author.
% Explanatory text should go in the []'s, 
% actual e-mail address or url should go in the {}'s for \email and \homepage.
% Please use the appropriate macro for the type of information

% \affiliation command applies to all authors since the last \affiliation command. 
% The \affiliation command should follow the other information.

\author{Bin Shao}

%\email[]{Your e-mail address}

%\homepage[]{Your web page}

%\thanks{}

%\altaffiliation{}

\affiliation{College of Information Technical Science, Nankai University, Tianjin 300071, China.}

\author{Min Feng}

%\email[]{Your e-mail address}

%\homepage[]{Your web page}

%\thanks{}

%\altaffiliation{}

\affiliation{School of Physics, Nankai University, Tianjin 300071, China.}

\author{Hong Liu}

%\email[]{Your e-mail address}

%\homepage[]{Your web page}

%\thanks{}

%\altaffiliation{}

\affiliation{Office of International Academic Exchanges, Nankai University, Tianjin 300071, China.}

\author{Jian Wu}

%\email[]{}

%\homepage[]{Your web page}

%\thanks{}

%\altaffiliation{}

\affiliation{Physics Department, Tsinghua University, Beijing 100084, China.}

% Collaboration name, if desired (requires use of superscriptaddress option in \documentclass). 

% \noaffiliation is required (may also be used with the \author command).

%\collaboration{}

%\noaffiliation

\author{Xu Zuo}

\email[Electronical Mail: ]{xzuo@nankai.edu.cn}

%\homepage[]{Your web page}

%\thanks{}

%\altaffiliation{}

\affiliation{College of Information Technical Science, Nankai University, Tianjin 300071, China.}

\date{\today}

\begin{abstract}

% insert abstract here

Based on first-principles calculation, it has been predicted that the magnetic anisotropy energy (MAE) in Co-doped ZnO (Co:ZnO) depends on electron-filling. Results show that the charge neutral Co:ZnO presents a \textquotedblleft easy plane\textquotedblright magnetic state. While modifying the total number of electrons, the easy axis rotates from in-plane to out-of-plane. The alternation of the MAE is considered to be the change of the ground state of Co ion, resulting from the relocating of electrons on Co \textit{d}-orbitals with electron-filling.

\end{abstract}

% insert suggested PACS numbers in braces on next line


\maketitle %\maketitle must follow title, authors, abstract and \pacs

% Body of paper goes here. Use proper sectioning commands. 
% References should be done using the \cite, \ref, and \label commands
%\section{}
%\label{}
%\subsection{}
%\subsubsection{}

Magnetic anisotropy energy (MAE) is a crucial parameter in high density magnetic storage \cite{809134, ross:3168}. A large magnitude of MAE will enhance the robustness of one bit to stabilize its magnetic spins. An efficient writing process requires a controllable direction of easy axis, which is determined by the sign of MAE. Co-doped ZnO (Co:ZnO), a possible diluted magnetic semiconductor (DMS) with high Curie temperature ($T_{c}$), has been demonstrated to be a strong single ion anisotropy material \cite{Sati}. Recently, the dependence of magnetic anisotropy on charge carrier concentration controlled by an external electric field has been observed in DMS \cite{chiba}. It offers a promising route to apply magnetic recording capability into information processing unit in current semiconductor devices. Therefore, it is highly desirable to understand the relation between MAE and carrier concentration in DMSs.

In this work, we carry out first-principles calculation based on density functional theory (DFT) to obtain MAE in Co:ZnO. Since adding the external electric-field will induce the accumulation of donors (electron/hole) \cite{Yamada2011}, the variation of carrier concentration is treated as electron-filling in calculations \cite{Wei}. Then, we investigate the dependence of MAE in Co:ZnO on electron-filling effect. 

A $ 2\times2\times2 $ supercell is created from the wurtzite primitive cell of ZnO with experimental lattice parameters and one Zn atom is substituted by Co ($\text{Co}_{0.0625}\text{Zn}_{0.9375}\text{O} $). With fixed lattice parameters, the internal coordinates are optimized by using Perdew-Burke-Ernzerhof (PBE) \cite{PBE} parameterization of generalized gradient approximation (GGA) as implemented in VASP package\cite{VASP}.

The electronic structure and magnetic properties are calculated after the structure optimization. We employ a corrected-band-gap scheme (CBGS) \cite{Shao}, enhancing the bandgap of pristine ZnO to 2.8 eV. Besides, the Hubbard \textit{U} (2 eV) is applied to Co \textit{d}-orbital \cite{Wei}. The cut-off energy of plane wave is 500 eV. For the integration in Brillouin zone, the tetrahedron method with Bl\"ochl corrections \cite{Blochl} is employed under a $5\times5\times3$ \textit{k}-mesh. The accuracy of electronic iterations is up to $10^{-6}$ eV. Then, spin-orbit coupling (SOC) is performed in non-self-consistent calculations based on the wavefunctions and charge distribution from self-consistent calculations without SOC. The MAE is estimated as $\Delta E=E_{[100]}-E_{[001]}$, where $E_{[100]}$ and $E_{[001]}$ are the total energy in [100] and [001] magnetization directions, respectively. Finally, the calculations above are repeated by adding extra electrons ($\delta N$) from -1.0 to 2.0 \textit{e} with respect to the initial value.

 \begin{table}%tableI
 	\caption{\label{TableI} The spin moment ($\text{M}_\text{S}$) and orbital moment ($\text{M}_\text{L}$)  (all in $\mu_\text{B}$) and MAE (meV) with electron-filling ($\delta N$), where $\delta N$ is the electron number added. The $\Delta\text{M}_\text{L}$ is calculated as $\text{M}^{[100]}_\text{L}-\text{M}^{[001]}_\text{L}$.} 
	\begin{ruledtabular}
 		\begin{tabular}{ccccccc}
 		&\multicolumn{2}{c}{$\text{M}_\text{S}$}&\multicolumn{2}{c}{$\text{M}_\text{L}$}\\ 
 		\cline{2-3}\cline{4-5}
		$\delta N$	&Total		&Co		&[001]	&[100]	&$\Delta\text{M}_\text{L}$	&MAE\\ 
		\hline
 		-1.0			&3.817		&2.959	&0.071	&0.067	&-0.004					&0.241\\
 		0			&2.857		&2.611	&0.082	&0.085	&0.003					&-0.107\\
 		1.0			&2.712		&2.509	&0.111	&0.093	&-0.018					&0.210\\
 		2.0			&2.449		&2.354	&0.126	&0.096	&-0.030					&0.319\\
 		\end{tabular}
 	\end{ruledtabular}
 \end{table}

\begin{figure}%figure1
 \includegraphics{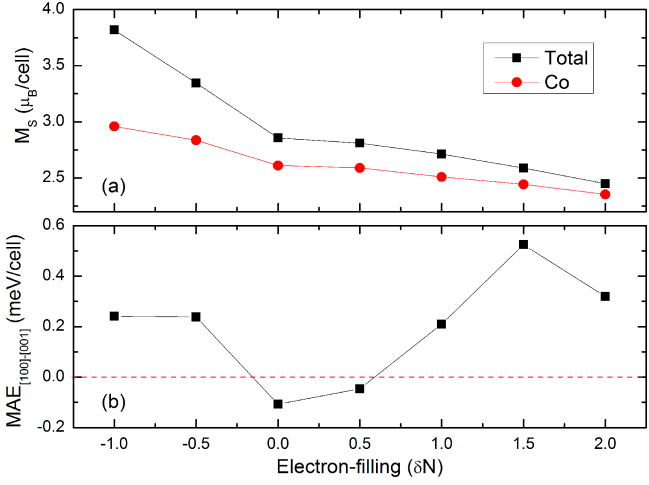}%
 \caption{\label{fig1}The dependence of (a) spin magnetic moment  and (b) MAE  on electron-filling. The positive MAE indicates that the easy axis parallels to the anatase \textit{c} axis.}%
\end{figure}

\begin{figure}%figure2
 \includegraphics{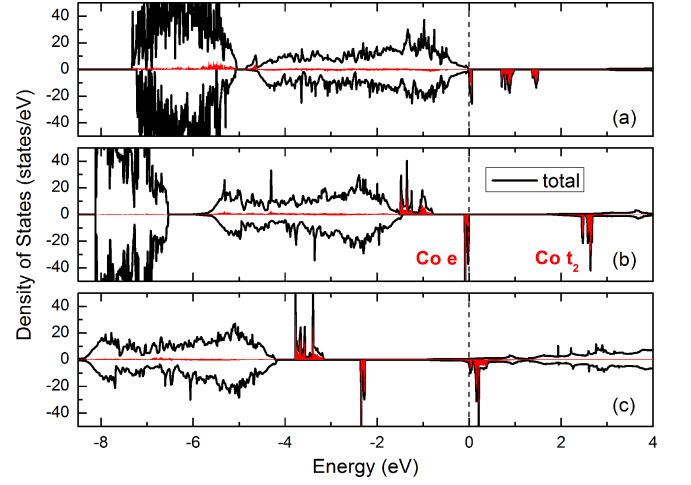}%
 \caption{\label{fig2}The density of states with CBGS and $U_{\text{Co}-d}$, (a) $\delta N=-1$, (b) $\delta N=0$ and (c) $\delta N=1$. The Co \textit{d}-orbitals are filled in red color. The vertical dash line refers to the Fermi level.}
\end{figure}

In Table~\ref{TableI}, we summarize the results of MAE, spin moment ($\text{M}_\text{S}$) and orbital moment ($\text{M}_\text{L}$) with electron-filling. Both of the total $\text{M}_\text{S}$ and the $\text{M}_\text{S}$ on Co drop monotonously when increasing the number of electrons (Fig.~\ref{fig1}). The MAE of charge neutral Co:ZnO is negative, implying an \textquotedblleft easy plane\textquotedblright magnetic state, which is in accordance with experimental results shown in Ref.~\onlinecite{Sati}. However, when increasing and decreasing one or more electrons, the sign of MAE alters, yielding the direction spin preferred switches from in-plane to out-of-plane. Meanwhile, the magnitude enlarges to more than 0.2 meV, even up to 0.319 meV for $\delta N=2$. Besides, it should be noted that the $\text{M}_\text{L}$ in the direction of preferential spin axis is larger than that in another, as given by Bruno's model \cite{Bruno1989}.
To interpret the variation of MAE with electron-filling, the electronic structure for different electron number is first checked. The density of states are calculated and projected to Co \textit{d}-orbital for $\delta N=1$, 0 and 1. For the neutral Co:ZnO [Fig.~\ref{fig2}(b)], the gap of pristine ZnO is preserved. The impurity bands, formed by the hybridization between the levels of the host Zn vacancy (oxygen dangling bonds) and Co \textit{d}-orbital, lie in the bandgap. The Co \textit{d}-orbital in majority spin, which is just above the valence band maximum (VBM) of the ZnO host, is completely occupied. The topmost occupied state is Co \textit{e} state in minority spin with 1.5 eV higher than VBM. Additionally, the $e-t_2$ split is about 2.3 eV, and the $t_2$ state in minority is pushed into conduction band (CB). The system is insulator. Our calculation confirms the electronic configuration of $\text{Co}^{2+}$, $d^7$, with a high spin state in neutral Co:ZnO, which is in agreement with previous work \cite{Wei}.

When decreasing one electron ($\delta N=-1$) [Fig.~\ref{fig2}(a)], the initial double degenerated and fully occupied \textit{e} state of Co in minority spin splits into two bands, one strongly hybrids with valence band of ZnO (O \textit{p}-orbital) and the other is empty, just locating above the Fermi level. For $\delta N=1$ [Fig.~\ref{fig2}(c)], doping with one electron, the empty $t_2$ state in minority spin is partially filled. The system becomes metallic. Compared to the neutral system, the electrons relocate in the \textit{d}-orbital of Co in minority spin when $\delta N=-1$ and 1, which finally results in the change of the ground state of Co ion.
\begin{figure}%figure3
 \includegraphics{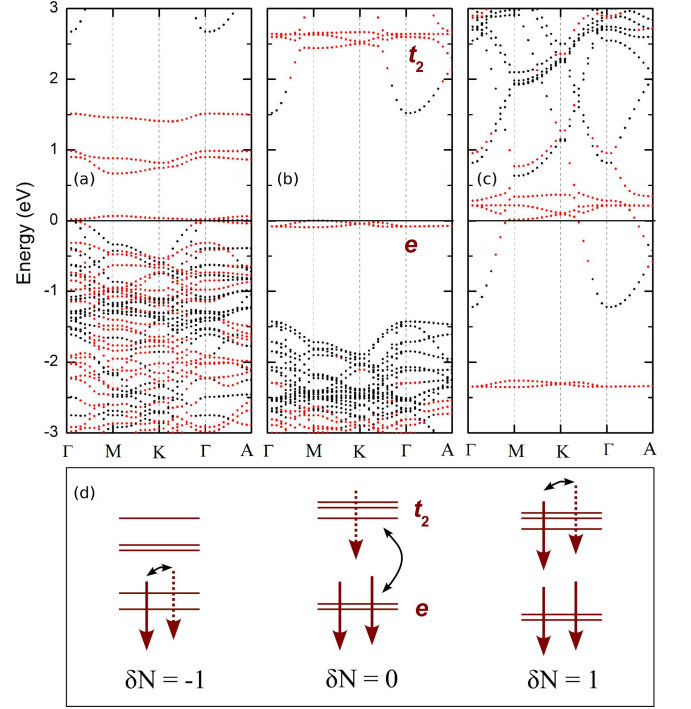}%
 \caption{\label{fig3}Bandstructure of minority spin along the high symmetric line in Brillouin zone, (a) $\delta N=-1$, (b) $\delta N=0$ and (c) $\delta N=1$. The red color of the point refers to the state with contribution from Co \textit{d}-orbital more than 2\%. The horizontal line at 0 eV is the Fermi level. (d) Schematic diagram (minority spin configuration) of the perturbation determining MAE in Co:ZnO with electron-filling. The MAE for $\delta N=0$ is decided by the perturbation between the \textit{e} level and $t_2$ level. However, that perturbation for $\delta N=-1$ and 1 exists within \textit{e} and $t_2$ manifold, respectively.}
\end{figure}

\begin{figure}%figure4
 \includegraphics{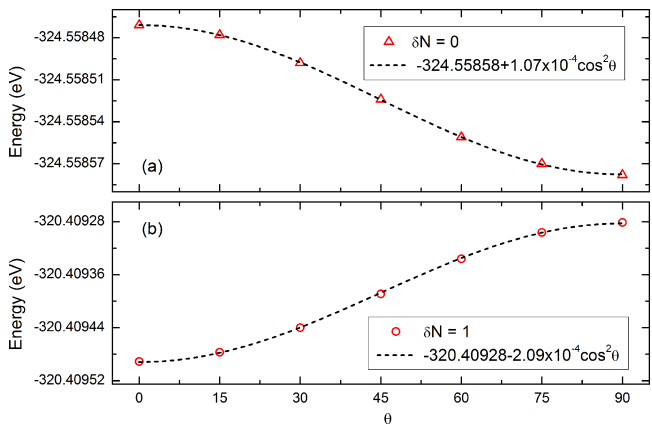}%
 \caption{\label{fig4}The dependence of the total energy on the angle ($\theta$) between magnetization direction and the wurtzite \textit{c} axis, (a) $\delta N=0$ and (b) $\delta N=1$.}
\end{figure}
As proposed by van Vleck \cite{PhysRev.52.1178}, the MAE is induced by the relativistic SOC interaction. Specifically, for 3d transition metal elements, the interaction is weaker than the crystal-field split. Therefore, according to the sing-ion anisotropy theory \cite{Yosida}, MAE can be estimated by the second-order perturbation between the ground state and excited states of magnetic ion. Since the Co ion exists as a high spin state in Co:ZnO (Fig.~\ref{fig2}), which implies the exchange split is larger than crystal-field effect, and the \textit{d}-orbital in majority spin is fully occupied, the variation of ground state of Co ion under electron-filling originates from the electron relocating in minority spin.

To describe this physical scenario straightforwardly, we calculate the bandstructure of minority spin for different $\delta N$ and at the same time mark the state with contribution from Co \textit{d}-orbital more than 2\%, as presented in Fig.~\ref{fig3}. For $\delta N=0$ [Fig.~\ref{fig3}(b)], the almost two-fold degenerated \textit{e} level in ground state is completely occupied, while the $t_2$ level in conduction band is empty. Therefore, the excited states associate with electrons hopping from the $e$ level and $t_2$ level, i.e., the MAE for charge neutral Co:ZnO is determined by the perturbation between $e$ level and $t_2$ level. However, as shown in Fig.~\ref{fig3}(a) and (b), this perturbation for $\delta N=-1$ and 1 exists within $e$ and $t_2$ manifold, respectively. We conclude that the change of the perturbation process [Fig.~\ref{fig3}(d)] finally results in the reorientation of easy axis with respect to $\delta N=0$.

Further, we calculate the total energy as a function of the angle ($\theta$) between the magnetization direction and the wurtzite \textit{c} axis. Obviously, as shown in Fig.~\ref{fig4}, we predict an easy-plane and easy-axis magnetic anisotropic state for $\delta N=0$ and 1, respectively. Moreover, the trend of the data implies that the contribution to MAE mainly comes from the second-order uniaxial anisotropy. By fitting the total energy to $C_0 +C_1 \cos ^2 \theta$ curve, the standard error for the two coefficients, $C_0$ and $C_1$, up to $10^{-6}$ supports the former speculation.

In conclusion, first-principles calculations are carried out to investigate the relationship between MAE and carrier concentration in Co:ZnO. Results show that Co:ZnO in neutral possesses an \textquotedblleft easy plane\textquotedblright magnetic state. While modifying the total number of electrons by one or more, the system alters the sign of MAE, switching into \textquotedblleft easy axis\textquotedblright magnetic state. The electronic configurations on Co ion with electron-filling are checked. We conclude that, due to the electron relocating on Co \textit{d}-orbital, the perturbation between the ground state and excited states for different total number of electrons has changed, finally reorienting the direction of spin axis. Further, we investigate the dependence of total energy on the magnetization direction. By fitting the data, a second-order uniaxial anisotropy has been found in Co:ZnO.

% If you have acknowledgments, this puts in the proper section head.
\begin{acknowledgments}
% Put your acknowledgments here.
This research was sponsored by National Natural Science Foundation of China (Grant No. 10970499), National Basic Research Program of China (973 Program, Grant No. 2011CB606405).
\end{acknowledgments}

% If in two-column mode, this environment will change to single-column format so that long equations can be displayed. 
% Use only when necessary.
%\begin{widetext}
%$$\mbox{put long equation here}$$
%\end{widetext}
% figures

% Create the reference section using BibTeX:
%\bibliography{manu}
%merlin.mbs aipnum4-1.bst 2010-07-25 4.21a (PWD, AO, DPC) hacked
%Control: key (0)
%Control: author (8) initials jnrlst
%Control: editor formatted (1) identically to author
%Control: production of article title (-1) disabled
%Control: page (0) single
%Control: year (1) truncated
%Control: production of eprint (0) enabled
%

\end{document}